\documentclass[11pt]{article}

\usepackage[final]{acl}

\usepackage{times}
\usepackage{latexsym}

\usepackage[T1]{fontenc}

\usepackage[utf8]{inputenc}

\usepackage{microtype}

\usepackage{inconsolata}

\usepackage{graphicx}

\usepackage{booktabs}
\graphicspath{{./images/}}

\usepackage[most]{tcolorbox}
\usepackage{tabularx}
\usepackage{booktabs}
\usepackage{array}

\title{Ranked by Position: Order Sensitivity as an Exploitable Attack Surface in LLM Listwise Recommenders}

\author{Ge Zhang \\
  Stanford University \\
  \texttt{gmzhang@stanford.edu} \\\And
  Jingru Cheng \\
  Stanford University \\
  \texttt{cjr63@stanford.edu} \\\And
  Huiyuan Chen \\
Independent Researcher \\
  \texttt{hxc501@case.edu} \\}

\begin{document}
\maketitle
\begin{abstract}
Large language models (LLMs) used as listwise rerankers in recommendation systems suffer from position bias when serializing candidate sets into prompts. We show this order sensitivity creates an exploitable attack surface: an attacker can promote a label-0 target into the top-$k$ solely by reordering candidates, without changing item content, labels, or model parameters. We introduce $\mathrm{promo}@k$ to quantify this vulnerability, measuring the fraction of label-0 targets that can be elevated into top-$k$ rankings via permutation. Evaluating across three domains (MovieLens, Amazon Books, and Amazon Fashion), $\mathrm{promo}@5$ reaches up to 0.57 at an attack budget of $R$ = 50 orderings. Furthermore, ordinary permutation stability predicts vulnerability without running the attack. While a bidirectional T5 encoder scorer reduces exposure, permutation-consistency regularization and architectural invariance effectively mitigate it. Pointwise scoring avoids the bias issue but degrades ranking quality. These results demonstrate that input candidate order in listwise LLM reranking is a security-relevant attack vector. Code and data are available at \url{https://github.com/geoz-lab/position_bias_attack}.
\end{abstract}

\begin{figure}[t]
  \centering
  \includegraphics[width=\columnwidth]{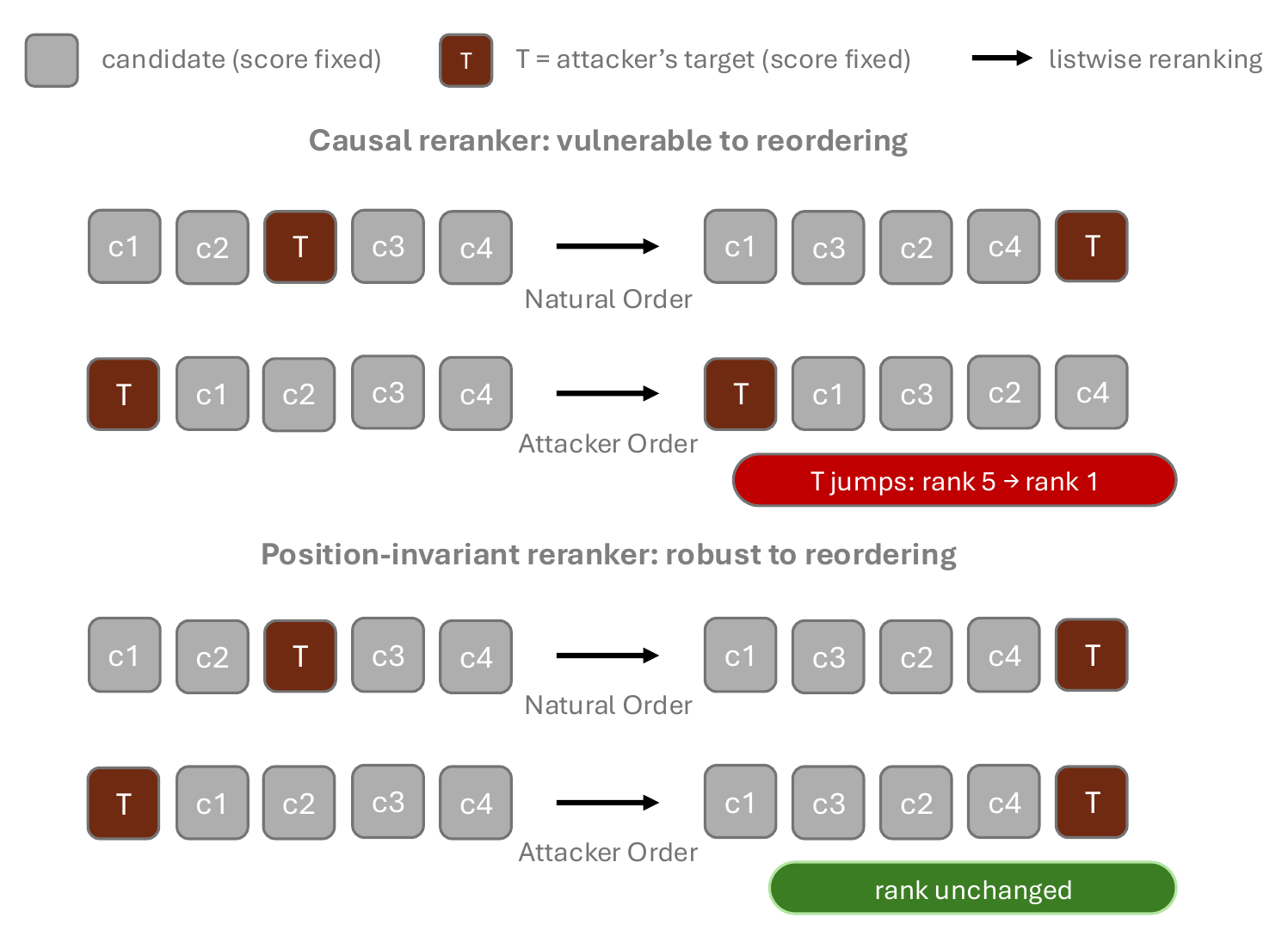}
  \caption{An ordering attack. The candidate set and its contents are fixed; only the \emph{input order} changes. A causal LLM reranker (top) produces a different output for the attacker's order, promoting a label-\(0\) target \(T\) to the top; a position-invariant reranker (bottom) is unaffected.}
  \label{fig:teaser}
\end{figure}

\section{Introduction}

Large language models (LLMs) are increasingly used as listwise rerankers in recommendation and retrieval systems. Given a user history or query and a candidate set, a listwise reranker serializes all candidates into one prompt, scores them jointly, and returns a ranked list~\cite{sun2023chatgpt,hou2024zeroshot,pradeep2023rankzephyr}. This formulation enables cross-candidate comparison but creates a semantic--structural mismatch: candidate items form an unordered set, whereas decoder-only LLMs process them as an ordered token sequence. Through causal attention and positional encodings such as RoPE~\cite{su2024roformer}, a candidate's score may therefore depend on its position even when its content and relevance are unchanged.

Order sensitivity has been observed in in-context learning and LLM-based evaluation~\cite{liu2024lostinthemiddle,wang2023notfair,shi2025judgingjudges}. In reranking, prior work mainly treats it as instability to be measured or mitigated, for example through permutation averaging or position-robust training~\cite{tang2024permsc,bito2026invarirank,bito2025rise}. We argue that this perspective understates the risk. Because candidate order is semantically arbitrary, systematic dependence on that order creates a manipulable input channel rather than merely random noise.

We therefore study candidate-order manipulation as an adversarial attack on listwise rerankers. The attacker changes only the presentation order of a fixed candidate set, without modifying item content, relevance labels, user behavior, or model parameters (Figure~\ref{fig:teaser}). In our threat model, the client supplies the candidate ordering and may try a limited number of permutations, retaining the one that gives a chosen target its best rank (Section~\ref{sec:setup}). We introduce $\mathrm{promo}@k$ to measure the fraction of label-0 targets moved from outside the top-$k$ into it through reordering alone. On the MovieLens anchor, trying 50 permutations manufactures top-5 exposure for 12\% of targets under a causal reranker, compared with 0\% under a position-invariant reference.

The vulnerability generalizes across eight causal rerankers and three recommendation domains. At an attack budget of $R=50$, $\mathrm{promo}@5$ reaches 0.57, with substantially larger exposure on the Amazon domains than on MovieLens. We find no evidence that exploitability decreases monotonically with model scale, while success increases smoothly with the attacker's permutation budget. Moreover, ordinary permutation stability is strongly associated with attack severity across 24 model--domain settings, even after controlling for domain identity, providing an attack-free signal for auditing order-manipulation risk (Section~\ref{sec:general}).

We make three contributions. First, we formalize candidate-order manipulation as an attack on listwise LLM rerankers and introduce \(\mathrm{promo}@k\) to measure manufactured top-k exposure. Second, we show that the vulnerability is structured, budget-dependent, and present across eight rerankers and three recommendation domains, with no evidence of a monotonic reduction with model scale. Third, we identify permutation stability as an attack-free audit signal and evaluate four defense families, finding that consistency regularization substantially reduces exposure and architectural invariance removes the tested attack channel.

\section{Related Work}

\textbf{LLM reranking and order sensitivity.}
LLMs have been widely studied as listwise rerankers for retrieval and recommendation, including RankGPT, open-source rerankers, and zero-shot recommendation methods~\cite{sun2023chatgpt,pradeep2023rankvicuna,pradeep2023rankzephyr,reddy2024first,hou2024zeroshot}. These methods exploit joint cross-candidate comparison but generally treat the presented candidate order as fixed. Meanwhile, LLM outputs are known to vary with input position in long-context reasoning, evaluation, recommendation, and other structured tasks ~\cite{liu2024lostinthemiddle,xiao2023attentionsinks,wang2023notfair,shi2025judgingjudges,li2024splitmerge,jinllm,bito2025rise,chen2024premise}. Prior work primarily frames such dependence as a capability, robustness, or evaluation issue. We instead study it as an attack surface. Because candidate sets are semantically unordered, an adversary may exploit their serialization order to manufacture top-$k$ exposure.

\textbf{Mitigating order sensitivity.}
Existing approaches reduce order dependence through architectural or encoding constraints~\cite{bito2026invarirank,yoon2024listt5,zhuang2024setwise,egressy2026set}, training-time regularization or calibration~\cite{chao2024alro,zhang2024positionaware,zhao2025simaug,qiao2026debiasfirst,joachims2017unbiased,ai2018unbiased,hager2024ultr,chen2025icr}, and inference-time aggregation~\cite{tang2024permsc,zeng2024rankfusion}.These methods are typically motivated by ranking quality, calibration, or stability. We evaluate representative defenses under an explicit order-manipulation threat model, treating permutation invariance as a security property. We also examine whether exploitability varies with model scale, rather than assuming that larger models are less position-sensitive.

\textbf{Ranking manipulation attacks.}
Prior attacks manipulate rankings through prompts, candidate content, ranking criteria, or user interactions~\cite{kumar2024sts,tang2025srp,liu2024attackchain,wang2024textsimu,ning2024cheatagent,chen2022denoising,qian2025jailbreak,anon2025raf}. Our attack is orthogonal to these approaches: it changes only the order of a fixed candidate set, while preserving item content, relevance labels, user behavior, and model parameters. This isolates candidate serialization as a distinct manipulation channel that prior content-based attacks often fix or control away.

\section{Problem Setup and Threat Model}
\label{sec:setup}

\textbf{Listwise reranking.}
Given a user context \(H\), such as an interaction history or search query, and a candidate set \(\mathcal{C}=\{c_1,\ldots,c_K\}\) with \(K=25\), a listwise reranker serializes the candidates according to an input order \(\pi\), scores them in a shared prompt, and returns the candidates sorted by their scores. For our decoder-only rerankers, the score of a candidate is the mean token log-probability over the candidate's text span. The temporal order of interactions inside \(H\) is preserved, as in standard sequential recommendation; \(\pi\) refers only to the presentation order of the candidate items. Since \(\mathcal{C}\) is a set once candidate content is fixed, permuting \(\pi\) should not change item relevance or the desired ranking, unless candidate order is explicitly intended as an input feature. We therefore treat dependence of the output ranking on \(\pi\) as unintended order sensitivity.

\textbf{Causal and position-invariant rerankers.}
Our main reranker is a decoder-only causal model with standard causal attention and RoPE positional embeddings. We compare it with a position-invariant reranker based on InvariRank~\cite{bito2026invarirank}, which uses a structured cross-candidate attention mask and shared candidate position IDs to remove dependence on candidate presentation order. The two rerankers use the same scoring function, loss, and training data; they differ only in how cross-candidate attention and candidate positions are represented. We use the position-invariant reranker as a reference defense, not as a new architecture proposed by this work.

\textbf{Permutation-consistency regularization.}
As a training-time defense, we add a consistency penalty that discourages the scorer from depending on candidate order. For each candidate set we sample \(P\) input orders \(\pi_1,\ldots,\pi_P\), score the set under each, and align the resulting score vectors back to canonical candidate identity. Writing \(p^{(\pi)}=\mathrm{softmax}(s^{(\pi)}/T)\) for the induced distribution over candidates, the penalty is the mean Kullback--Leibler divergence from the first order to each remaining one,
\begin{equation}
\mathcal{L}_{\mathrm{perm}}
=
\frac{1}{P-1}\sum_{j=2}^{P}
\mathrm{KL}\!\big(p^{(\pi_1)} \,\|\, p^{(\pi_j)}\big),
\end{equation}
and the training objective is 
\(\mathcal{L}=\mathcal{L}_{\mathrm{rank}}+\lambda_{\mathrm{perm}}\mathcal{L}_{\mathrm{perm}}\),
where \(\mathcal{L}_{\mathrm{rank}}\) is the LambdaRank loss averaged over the same \(P\) orders. We use \(P=2\), \(T=1\), and \(\lambda_{\mathrm{perm}}=1\). Gradients propagate through both branches: \(\pi_1\) serves as the reference order but is not treated as a detached target, so the penalty pulls the two score distributions toward each other rather than distilling one into the other. The defense is approximate rather than exact, nothing in the architecture forbids order dependence at inference, which is why a small residual surface survives (Table~\ref{tab:defense_landscape}).

\textbf{Candidate labels.} Candidate labels are derived from temporally held-out future interactions: observed ratings map to graded relevance labels \(1\)--\(4\), while LightGCN-retrieved negatives and catalog filler items receive label \(0\). Label-\(0\) candidates thus have no meaningful internal ground-truth order, and our attack metrics do not assume one is more relevant than another; full label rules and per-domain configurations are in Appendix~\ref{app:dataset}.

\textbf{Threat model.}
The attacker aims to promote a chosen target item \(c^\ast\) into the top-\(k\) results by manipulating only the candidate presentation order. In our main evaluation, \(c^\ast\) is sampled from label-\(0\) candidates, so the attack measures manufactured exposure for items that are label-0 under the held-out interaction labels. The attacker may control or influence the input order \(\pi\), but cannot modify item content, relevance labels, user behavior, or model parameters.

We focus on a shared or API-based reranking service in which the client supplies an ordered candidate list and observes the returned ranking. The attacker is budget-limited: for each target, it may try at most \(R\) candidate orderings and keep the ordering that gives the target its best rank. This models a black-box ordering attack that uses only ranking outputs.

\textbf{Exploitability metrics.}
For an evaluation instance \(i\), let \(c_i^\ast\) denote the target item and let
\(r_{i,\pi}(c_i^\ast)\) be the output rank of the target under candidate ordering
\(\pi\), where rank \(0\) is best. We draw a pool of \(R_{\max}\) random orderings
\(\Pi_{R_{\max}}=(\pi_1,\ldots,\pi_{R_{\max}})\) and write
\(\Pi_R=(\pi_1,\ldots,\pi_R)\) for its first \(R\) elements, so that
\(\Pi_1\subseteq\cdots\subseteq\Pi_{R_{\max}}\). The attacker's budget is \(R\);
the pool size \(R_{\max}\) is a property of the measurement rather than of the attacker.

We summarize the target's typical, attack-free behavior by its mean rank over the
full pool,
\begin{equation}
    \bar r_i(c_i^\ast)
    =
    \frac{1}{R_{\max}}
    \sum_{\pi\in\Pi_{R_{\max}}}
    r_{i,\pi}(c_i^\ast),
\end{equation}
and define its adversarial rank at budget \(R\) as the best rank achieved by any
ordering the attacker tried,
\begin{equation}
    r_i^{\mathrm{adv}}(c_i^\ast)
    =
    \min_{\pi\in\Pi_R}
    r_{i,\pi}(c_i^\ast).
\end{equation}
For brevity we write \(\bar r_i\) and \(r_i^{\mathrm{adv}}\) when the target is clear from context. Fixing \(\bar r_i\) at the full pool keeps the promotion criterion below independent of \(R\), which makes \(\mathrm{promo}@k\) monotonically non-decreasing in the attacker's budget. Unless otherwise noted we use \(R_{\max}=R=50\).

We report \textbf{rank gain}, the number of positions the target gains under attack (\(\bar r_i - r_i^{\mathrm{adv}}\)), and \(\mathbf{\mathrm{Into}@5}\), the fraction of targets reaching the top five under at least one attempted ordering (\(r_i^{\mathrm{adv}} < 5\)), as a supporting diagnostic. Unlike \(\mathrm{promo}@5\) below, \(\mathrm{Into}@5\) does not require the target to start outside the top five.

Third, \(\mathrm{promo}@k\) measures newly manufactured top-\(k\) exposure. Over the \(N\) evaluation instances we define
\begin{equation}
\mathrm{promo}@k
=
\frac{1}{N}\sum_{i=1}^{N}
\mathbf{1}\big[\, \bar r_i \ge k > r_i^{\mathrm{adv}} \,\big],
\end{equation}
the fraction of all targets that lie outside the top-\(k\) under attack-free ordering \emph{and} can be driven inside it by reordering alone. Unlike \(\mathrm{Into}@k\), it excludes targets already ranked inside the top-\(k\) without an attack. Because the denominator is the full evaluation set rather than an eligibility-filtered subset, \(\mathrm{promo}@k\) is a population-level exposure rate and is directly comparable across models and domains.

\textbf{Permutation stability.}
To audit order sensitivity without running the adversarial search, we measure ordinary permutation stability. For each candidate set, we sample random candidate permutations, obtain one output ranking per permutation, and compute Kendall's \(\tau\) between pairs of output rankings. We average \(\tau\) over permutation pairs and candidate sets. Higher \(\tau\) means that the reranker produces more stable rankings under candidate reshuffling; lower \(\tau\) indicates stronger dependence on presentation order.

\textbf{Experimental setup.}
We fine tune rerankers with LoRA (\(r=16\), \(\alpha=32\)) for 500 steps, using sequence length 4096 and LambdaRank loss. Experiments cover three domains: MovieLens-32M~\cite{harper2015movielens} and two Amazon Reviews categories~\cite{hou2024amazonreviews}, Books and Fashion. Each candidate set contains \(K=25\) items. We initialize candidate sets with LightGCN~\cite{he2020lightgcn} first-stage retrieval, insert the selected held-out future items used to define relevance labels, and add catalog filler items when needed to keep the candidate-set size fixed. This insertion step is important for the sparse Amazon domains, where LightGCN recall@100 is only about \(1\%\).

For MovieLens and Amazon Books, we use a history window of 20 interactions. For Amazon Fashion, we use a shorter history window of 5 interactions to retain enough eligible users under its sparser interaction histories. Our anchor model is Llama-3.2-3B, evaluated with three seeds on MovieLens. We additionally sweep causal-only Qwen3 models from 0.6B to 14B, Llama-3.1-8B, and Mistral-7B across all three domains. We also include a bidirectional T5 encoder scorer as a mechanism comparison. Following the encoder-based design of ListT5~\cite{yoon2024listt5}, it encodes the serialized candidate list bidirectionally, but it is trained by us rather than taken from released weights: a flan-t5-base backbone with LoRA (\(r=32\)) for 2{,}000 steps at learning rate \(10^{-3}\), scoring candidates from pooled hidden states under a softmax cross-entropy ranking loss. We therefore describe it as ListT5-style rather than as ListT5 itself.

Evaluation sample sizes depend on the metric. Standard ranking effectiveness and permutation-stability metrics, including nDCG@10 and Kendall's \(\tau\), use 2{,}000 candidate sets. Position-sensitivity and headline attack analyses use 300 candidate sets; headline attack results are reported at budget \(R=50\), and the budget curve evaluates \(R\in\{1,5,10,20,50\}\). For the attack analyses we sample one label-\(0\) target per candidate set, giving
\(N=300\) evaluation instances per model--domain cell. Additional baselines and defense ablations, including pointwise scoring and test-time order averaging, are reported in the corresponding tables and appendix.

All anchor attack metrics---the mechanism comparison, defense ablations, prompt ablations, and the attack-budget curve---are computed from a single reference run with \(R_{\max}=50\) on the MovieLens anchor (seed 42), so that values are directly comparable across tables. 

\section{Position Bias Is Structured}
\label{sec:structure}

We first ask whether candidate-position effects are systematic or merely random noise. For a fixed target candidate \(c^\ast\), we keep its content and relevance label unchanged, move it through each input position \(0,\ldots,K-1\), keep the remaining candidate set fixed, and record the target's output rank. We aggregate results over candidate sets and stratify by relevance label.

\begin{figure}[t]
  \centering
  \includegraphics[width=0.9\columnwidth]{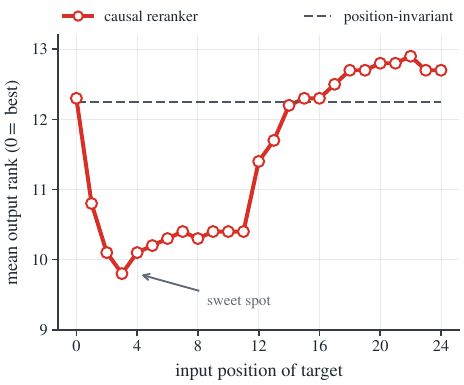}
    \caption{\textbf{Position bias is structured.} Mean output rank of a fixed target candidate as its input position is varied. The causal reranker exhibits a clear positional sweet spot, while the position-invariant reranker is nearly flat.}
  \label{fig:curve}
\end{figure}

Figure~\ref{fig:curve} shows a structured position-effect curve on the MovieLens anchor. The causal reranker exhibits a clear positional sweet spot: label-\(0\) targets receive better output ranks in early-middle input positions, while the first and tail positions are worse. By contrast, the position-invariant reranker is nearly flat. We summarize this effect using the position-curve range, defined as the difference between the worst and best mean output rank across input positions. For label-\(0\) targets, the range is \(3.09\) for the causal reranker but only \(0.06\) for the position-invariant reranker. The resulting positional sweet spots make candidate order predictable enough to search adversarially.


\section{Exploitability}
\label{sec:exploit}

\textbf{Attacker.}
We use budget-limited random-order search. For each target \(c^\ast\), the attacker samples \(R\) candidate orderings and keeps the ordering that gives \(c^\ast\) its best output rank. Unless otherwise noted, headline attack results use \(R=50\). We use black-box random search over at most $R$ permutations and retain the ordering yielding the target’s best rank.

\begin{figure}[t]
  \centering
  \includegraphics[width=0.9\columnwidth]{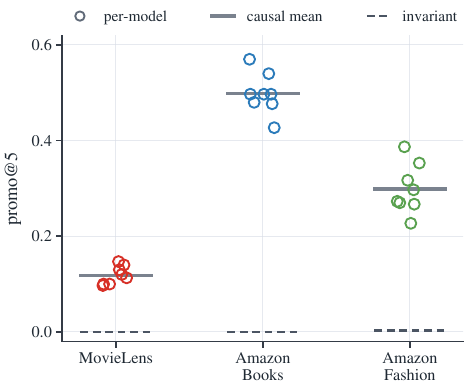}
  \caption{
  \textbf{Manufactured exposure across domains.} Each point is one causal model; horizontal bars denote domain means; dashed lines show the position-invariant reference.
  }
  \label{fig:promo}
\end{figure}

Figure~\ref{fig:promo} reports the main exploitability result using \(\mathrm{promo}@5\). On the MovieLens anchor, the causal reranker is vulnerable to order manipulation: changing only candidate order yields a rank gain of \(3.34\) positions, moves the target into the top five in \(27.3\%\) of cases, and manufactures new top-5 exposure for \(12.0\%\) of label-\(0\) targets. By contrast, the position-invariant reranker has \(\mathrm{promo}@5=0.000\). Thus, for the causal reranker, input order alone can create top-5 exposure for label-\(0\) targets that would otherwise remain outside the top five.

The attack surface is substantially larger on the Amazon domains (Figure~\ref{fig:promo}); the position-invariant reference stays at or near zero throughout. Weaker content signals appear to make candidate position a stronger lever.

\section{Mechanism Comparisons and Defenses}
\label{sec:defense}

\paragraph{Encoder and pointwise controls.}
Table~\ref{tab:mechanism_comparison} compares the causal reranker with three controls on the MovieLens anchor. The T5 encoder scorer encodes the serialized candidate list bidirectionally rather than with decoder-only causal attention, but it still uses candidate order and positional information. Pointwise scoring removes candidate-order dependence by scoring each candidate independently, but gives up listwise comparison and requires \(O(K)\) scoring.

The T5 encoder scorer improves ranking quality and reduces order exploitability, but does not eliminate it: \(\mathrm{promo}@5\) drops from \(0.120\) to \(0.057\), and the label-\(0\) position-curve range drops from \(3.09\) to \(0.71\). Architectural invariance and pointwise scoring remove manufactured exposure, with \(\mathrm{promo}@5=0\). The pointwise reference still has \(\mathrm{Into}@5=0.157\), meaning that some label-\(0\) targets can receive top-5 exposure because of fixed content-based model errors; however, its \(\mathrm{promo}@5=0\) and zero rank gain show that this exposure is not manufactured by candidate-order manipulation. Because the T5 scorer also differs in backbone, loss, and scoring head, we interpret it as a mechanism comparison rather than evidence that bidirectionality alone causes the reduction.

\begin{table*}[t]
\centering
\caption{
\textbf{Mechanism comparison on the MovieLens anchor.} Range denotes the label-\(0\) position-curve range.
}
\label{tab:mechanism_comparison}
\small
\begin{tabular}{llcccccc}
\toprule
Model & Structure & nDCG@10 & Range & \(\tau\) & Gain & Into@5 & \(\mathrm{promo}@5\) \\
\midrule
Causal Llama-3B
& causal decoder
& 0.829 & 3.09 & 0.845 & 3.34 & 0.273 & 0.120 \\

T5 encoder scorer
& bidirectional encoder
& \textbf{0.876} & 0.71 & 0.913 & 1.88 & 0.197 & 0.057 \\

Architectural invariance
& constrained listwise
& 0.784 & 0.06 & 0.992 & 0.22 & 0.157 & \textbf{0.000} \\

Pointwise reference
& isolated scoring
& 0.763 & \textbf{0.00} & \textbf{1.000} & \textbf{0.00} & 0.157 & \textbf{0.000} \\
\bottomrule
\end{tabular}
\end{table*}

\paragraph{Defense results.}
Table~\ref{tab:defense_landscape} summarizes the main defense results across domains. Naive order augmentation is not sufficient: on the MovieLens anchor, training on permuted candidate orders leaves \(\mathrm{promo}@5\) nearly unchanged at \(0.117\) as shown in Table~\ref{tab:anchor_defense_full}, compared with the baseline value of \(0.120\). In contrast, permutation-consistency regularization substantially reduces the attack surface, lowering \(\mathrm{promo}@5\) from \(0.120\) to \(0.007\) on MovieLens, from \(0.480\) to \(0.067\) on Books, and from \(0.297\) to \(0.047\) on Fashion. Exact architectural invariance closes the residual surface, with \(\mathrm{promo}@5=0.000\) on MovieLens and Books and \(0.003\) on Fashion. 


The quality--robustness trade-off is domain-dependent: on Books, exact invariance zeroes \(\mathrm{promo}@5\) at the largest nDCG@10 cost, while permutation-consistency regularization retains more quality for a still-large reduction (Table~\ref{tab:defense_landscape}). Costs are smaller on MovieLens and Fashion.

\begin{table*}[t]
\centering
\caption{
\textbf{Cross-domain defense landscape.} Lower \(\mathrm{promo}@5\) is better. Exact indicates whether the method is invariant by construction.}
\label{tab:defense_landscape}
\small
\begin{tabular}{lccccccc}
\toprule
& \multicolumn{2}{c}{MovieLens}
& \multicolumn{2}{c}{Amazon Books}
& \multicolumn{2}{c}{Amazon Fashion}
& \\
\cmidrule(lr){2-3}
\cmidrule(lr){4-5}
\cmidrule(lr){6-7}
Defense / model
& nDCG@10 & \(\mathrm{promo}@5\)
& nDCG@10 & \(\mathrm{promo}@5\)
& nDCG@10 & \(\mathrm{promo}@5\)
& Exact \\
\midrule
Baseline causal
& 0.829 & 0.120
& 0.618 & 0.480
& 0.449 & 0.297
& No \\

Architectural invariance
& 0.784 & \textbf{0.000}
& 0.510 & \textbf{0.000}
& 0.424 & \textbf{0.003}
& Yes \\

Permutation-consistency KL
& 0.777 & 0.007
& 0.552 & 0.067
& 0.424 & 0.047
& Approx. \\

Pointwise reference
& 0.763 & \textbf{0.000}
& -- & \textbf{0.000}
& -- & \textbf{0.000}
& Yes \\
\bottomrule
\end{tabular}

\vspace{0.25em}
\footnotesize{
Pointwise nDCG is measured on the MovieLens anchor only. Its cross-domain \(\mathrm{promo}@5=0\) follows by construction because candidates are scored independently.
}
\end{table*}

\section{Generalization}
\label{sec:general}

We evaluate whether order exploitability generalizes beyond the MovieLens anchor using eight causal rerankers across MovieLens-32M, Amazon Books, and Amazon Fashion. The sweep includes Qwen3 models from 0.6B to 14B parameters, Llama-3.1-8B, Mistral-7B, and Llama-3.2-3B. Unless noted otherwise, we report \(\mathrm{promo}@5\) at \(R=50\); full results appear in Appendix Table~\ref{tab:cross_model_domain_full}.

\begin{figure}[t]
  \centering
  \includegraphics[width=0.9\columnwidth]{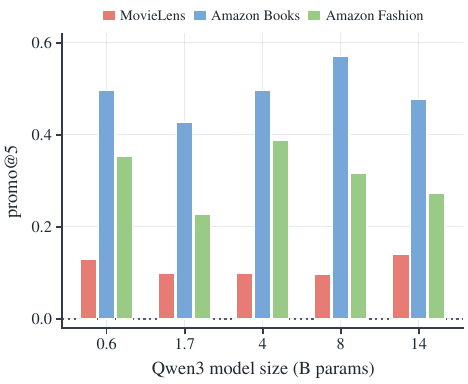}
  \caption{\textbf{Vulnerability does not monotonically decrease with model size.}}
  \label{fig:scale}
\end{figure}

\paragraph{Model scale and domain severity.}
Figure~\ref{fig:scale} shows no monotonic decrease in exploitability with model size. Across the Qwen3 sweep, \(\mathrm{promo}@5\) ranges from \(0.10\)--\(0.14\) on MovieLens, \(0.23\)--\(0.39\) on Amazon Fashion, and \(0.43\)--\(0.57\) on Amazon Books. Seed variation on the MovieLens anchor is comparable to or larger than the differences among backbones, so small within-domain gaps should not be over-interpreted. The much larger cross-domain differences nevertheless indicate that domain characteristics influence exploitability more strongly than parameter count.

This domain pattern is consistent with weaker content signals in the Amazon datasets. MovieLens provides structured metadata, whereas the Amazon domains are sparser and have LightGCN Recall@100 of about \(0.013\). This association suggests that position may become a stronger cue when content evidence is weak, but it does not establish causality. The position-invariant reference remains at or near zero across domains, although its ranking-quality cost is largest on Books (Table~\ref{tab:defense_landscape}).

\paragraph{Permutation stability as an audit signal.}
To test whether vulnerability can be audited without adversarial search, we correlate Kendall's \(\tau\) under random candidate reshuffling with \(\mathrm{promo}@5\) across the 24 causal model--domain cells.

\begin{figure}[t]
  \centering
  \includegraphics[width=0.9\columnwidth]{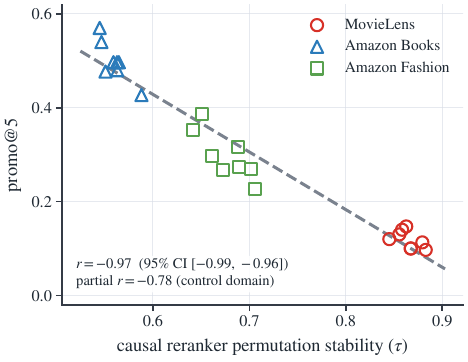}
  \caption{\textbf{Permutation stability tracks order exploitability.} Kendall's \(\tau\) versus \(\mathrm{promo}@5\) across 24 causal model--domain cells. Lower stability is associated with greater manufactured exposure.}
  \label{fig:corr}
\end{figure}

Figure~\ref{fig:corr} shows a strong negative association between permutation stability and exploitability (\(r=-0.974\), 95\% bootstrap CI \([-0.989,-0.956]\)). The relationship remains substantial after controlling for domain identity (\(r=-0.776\), \(p=8\times10^{-6}\)), indicating that stability is not merely a proxy for domain difficulty. Permutation stability therefore provides a practical, attack-free audit signal for order exploitability.

\paragraph{Attack budget sensitivity.}
We also vary the attack budget on the MovieLens anchor. For the causal reranker, \(\mathrm{promo}@5\) rises from \(0.023\) at \(R=1\) to \(0.120\) at \(R=50\); the T5 encoder follows the same trend at lower magnitude, reaching \(0.057\). The position-invariant and pointwise references remain at zero, showing that the budget effect is specific to order-sensitive scoring.

\begin{figure}[t]
  \centering
  \includegraphics[width=0.9\columnwidth]{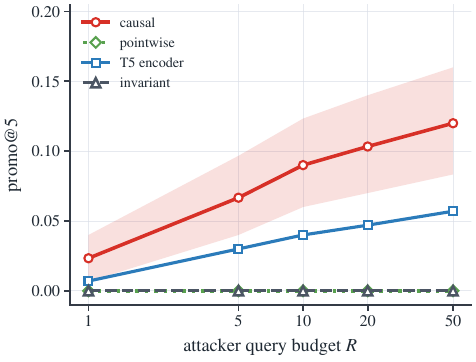}
  \caption{\textbf{Attack success increases with query budget.} \(\mathrm{promo}@5\) grows with the number of attempted orderings for order-sensitive scorers, while the position-invariant and pointwise references remain at zero. Shading indicates 95\% bootstrap confidence intervals.}
  \label{fig:budget}
\end{figure}



\section{Conclusion}

Candidate order is a manipulable attack surface to listwise LLM rerankers. Across models and domains, reordering alone promotes label-0 targets, with greater exposure under larger query budgets and less stable rerankers. Permutation stability therefore provides a practical audit signal. Consistency regularization mitigates most exposure, while architectural invariance removes the tested attack channel, subject to a domain-dependent ranking-quality cost. Listwise rerankers should treat permutation robustness as a security requirement rather than only a stability property. The same audit principle may apply to other systems that serialize semantically unordered alternatives, although this remains to be tested.



\section*{Limitations}
First, our threat model assumes the adversary can manipulate candidate presentation order, which aligns with API-based or shared reranking services but less so with systems using fixed or hidden candidate assembly. By evaluating single-stage reranking without altering retrieval pools, item content, or model parameters, we isolate candidate order as an attack channel, though we do not capture complex multi-stage attacks. Additionally, because label-$0$ targets are defined relative to held-out interaction data, $\mathrm{promo}@k$ measures manufactured exposure for non-positive evaluation items rather than intrinsic user irrelevance. Furthermore, our budget-limited random search provides a conservative lower bound on exploitability, as adaptive or optimized search strategies could uncover more effective orderings.

Second, our experimental scope and analytical claims carry specific boundaries. While primary defense families are evaluated across domains, full architectural and training ablations are centered on our anchor setting; additionally, pointwise scoring serves as an invariant baseline rather than a direct drop-in replacement, given its distinct computational profile. Finally, the observed relationship between Kendall's $\tau$ and $\mathrm{promo}@5$ across our 24 model--domain settings is observational rather than causal. Consequently, we position permutation stability as a practical, attack-free auditing signal rather than a proven causal mechanism of exploitability.

\section*{Ethics Statement}

This work is intended to support auditing and mitigation of order-based vulnerabilities in LLM rerankers. We report aggregate vulnerability magnitudes, evaluate defenses, and avoid releasing an optimized order-manipulation pipeline. The attack studied here changes only candidate order and is therefore useful for isolating the positional channel; it should not be interpreted as guidance for manipulating deployed ranking systems. The content-edit diagnostic, if used, is limited to separating positional effects from content effects and is not designed as a competitive content attack. By identifying effective mitigations such as architectural invariance and permutation-consistency regularization, our goal is to help practitioners detect and reduce order-manipulation risk.

\bibliography{custom}

@inproceedings{jinllm,
  title={LLM Maybe LongLM: SelfExtend LLM Context Window Without Tuning},
  author={Jin, Hongye and Han, Xiaotian and Yang, Jingfeng and Jiang, Zhimeng and Liu, Zirui and Chang, Chia-Yuan and Chen, Huiyuan and Hu, Xia},
  booktitle={Forty-first International Conference on Machine Learning},
  year={2024}
}

@inproceedings{bito2026invarirank,
  title     = {One Pass, Any Order: Position-Invariant Listwise Reranking for LLM-Based Recommendation},
  author    = {Bito, Ethan and Ren, Yongli and He, Estrid},
  booktitle = {Proceedings of the 49th International ACM SIGIR Conference on Research and Development in Information Retrieval},
  year      = {2026},
  publisher = {ACM}}

@article{bito2025rise,
  title   = {Evaluating Position Bias in Large Language Model Recommendations},
  author  = {Bito, Ethan and Ren, Yongli and He, Estrid},
  journal = {arXiv preprint arXiv:2508.02020},
  year    = {2025}
}

@inproceedings{qiao2026debiasfirst,
  title     = {LLM-based Listwise Reranking under the Effect of Positional Bias},
  author    = {Qiao, Jingfen and Huang, Jin and Ma, Xinyu and Wang, Shuaiqiang and Yin, Dawei and Kanoulas, Evangelos and Yates, Andrew},
  booktitle = {Advances in Information Retrieval},
  series    = {Lecture Notes in Computer Science},
  publisher = {Springer},
  year      = {2026}
}

@article{pradeep2023rankvicuna,
  title   = {RankVicuna: Zero-Shot Listwise Document Reranking with Open-Source Large Language Models},
  author  = {Pradeep, Ronak and Sharifymoghaddam, Sahel and Lin, Jimmy},
  journal = {arXiv preprint arXiv:2309.15088},
  year    = {2023}
}

@article{pradeep2023rankzephyr,
  title   = {RankZephyr: Effective and Robust Zero-Shot Listwise Reranking is a Breeze!},
  author  = {Pradeep, Ronak and Sharifymoghaddam, Sahel and Lin, Jimmy},
  journal = {arXiv preprint arXiv:2312.02724},
  year    = {2023}
}

@article{reddy2024first,
  title   = {FIRST: Faster Improved Listwise Reranking with Single Token Decoding},
  author  = {Reddy, Revanth Gangi and Doo, JaeHyeok and Xu, Yifei and Sultan, Md Arafat and Swain, Deevya and Sil, Avirup and Ji, Heng},
  journal = {arXiv preprint arXiv:2406.15657},
  year    = {2024}
}

@article{yoon2024listt5,
  title   = {ListT5: Listwise Reranking with Fusion-in-Decoder Improves Zero-shot Retrieval},
  author  = {Yoon, Soyoung and Choi, Eunbi and Kim, Jiyeon and Yun, Hyeongu and Kim, Yireun and Hwang, Seung-won},
  journal = {arXiv preprint arXiv:2402.15838},
  year    = {2024}
}

@inproceedings{chao2024alro,
  title     = {Make Large Language Model a Better Ranker},
  author    = {Chao, Wen-Shuo and Zheng, Zhi and Zhu, Hengshu and Liu, Hao},
  booktitle = {Findings of the Association for Computational Linguistics: EMNLP 2024},
  year      = {2024}
}

@article{zhang2024positionaware,
  title   = {Position-Aware Parameter-Efficient Fine-Tuning Approach for Reducing Positional Bias in LLMs},
  author  = {Zhang, Zheng and Yang, Fan and Jiang, Ziyan and Chen, Zheng and Zhao, Zhengyang and Ma, Chengyuan and Zhao, Liang and Liu, Yang},
  journal = {arXiv preprint arXiv:2404.01430},
  year    = {2024}
}

@inproceedings{tang2024permsc,
  title     = {Found in the Middle: Permutation Self-Consistency Improves Listwise Ranking in Large Language Models},
  author    = {Tang, Raphael and Zhang, Crystina and Ma, Xueguang and Lin, Jimmy and Ture, Ferhan},
  booktitle = {Proceedings of the 2024 Conference of the North American Chapter of the Association for Computational Linguistics (NAACL 2024)},
  year      = {2024}
}

@article{zeng2024rankfusion,
  title   = {LLM-RankFusion: Mitigating Intrinsic Inconsistency in LLM-based Ranking},
  author  = {Zeng, Yifan and Tendolkar, Ojas and Baartmans, Raymond and Wu, Qingyun and Chen, Lizhong and Wang, Huazheng},
  journal = {arXiv preprint arXiv:2406.00231},
  year    = {2024}
}

@inproceedings{zhuang2024setwise,
  title     = {A Setwise Approach for Effective and Highly Efficient Zero-shot Ranking with Large Language Models},
  author    = {Zhuang, Shengyao and Zhuang, Honglei and Koopman, Bevan and Zuccon, Guido},
  booktitle = {Proceedings of the 47th International ACM SIGIR Conference on Research and Development in Information Retrieval},
  year      = {2024}}

@inproceedings{chen2025icr,
  title     = {Attention in Large Language Models Yields Efficient Zero-Shot Re-Rankers},
  author    = {Chen, Shijie and Guti{\'e}rrez, Bernal Jim{\'e}nez and Su, Yu},
  booktitle = {Proceedings of the International Conference on Learning Representations (ICLR 2025)},
  year      = {2025}
}

@article{kumar2024sts,
  title   = {Manipulating Large Language Models to Increase Product Visibility},
  author  = {Kumar, Sachin and Lakkaraju, Himabindu},
  journal = {arXiv preprint arXiv:2404.07981},
  year    = {2024}
}

@article{tang2025srp,
  title   = {StealthRank: LLM Ranking Manipulation via Stealthy Prompt Optimization},
  author  = {Tang, Yiming and others},
  journal = {arXiv preprint arXiv:2504.05804},
  year    = {2025}
}

@inproceedings{anon2025raf,
  title={Are LLMs Reliable Rankers? Rank Manipulation via Two-Stage Token Optimization},
  author={Xing, Tiancheng and Li, Jerry and Du, Yixuan and Hu, Xiyang},
  booktitle={Proceedings of the 64th Annual Meeting of the Association for Computational Linguistics (Volume 1: Long Papers)},
  pages={9120--9132},
  year={2026}
}

@article{qian2025jailbreak,
  title   = {Jailbreaking LLM-based Rankers via Task Substitution and Ranking Criterion Manipulation},
  author  = {Qian, Hongjin and others},
  journal = {arXiv preprint arXiv:2510.10003},
  year    = {2025}
}

@techreport{burges2010ranknet,
  title     = {From {RankNet} to {LambdaRank} to {LambdaMART}: An Overview},
  author    = {Burges, Christopher J.C.},
  year      = {2010},
  number    = {MSR-TR-2010-82},
  institution = {Microsoft Research},
}

@inproceedings{wang2018lambdaloss,
  title     = {The {LambdaLoss} Framework for Ranking Metric Optimization},
  author    = {Wang, Xuanhui and Li, Cheng and Golbandi, Nadav and Bendersky, Michael and Najork, Marc},
  booktitle = {Proceedings of the 27th ACM International Conference on Information and Knowledge Management (CIKM)},
  year      = {2018},
  pages     = {1313--1322}
}

@inproceedings{liu2024attackchain,
  title     = {Attack-in-the-Chain: Bootstrapping Large Language Models for Attacks Against Black-box Neural Ranking Models},
  author    = {Liu, Yu-An and others},
  booktitle = {Proceedings of the AAAI Conference on Artificial Intelligence},
  year      = {2024}
}

@article{wang2024textsimu,
  title   = {TextSimu: LLM-driven N-persona Simulation Attack for Recommendation Exposure},
  author  = {Wang, Zongwei and others},
  journal = {arXiv preprint arXiv:2409.10001},
  year    = {2024}
}

@article{ning2024cheatagent,
  title   = {CheatAgent: Attacking Recommender Systems via LLM Agent},
  author  = {Ning, Liang-bo and others},
  journal = {arXiv preprint arXiv:2409.10002},
  year    = {2024}
}

@article{liu2024lostinthemiddle,
  title   = {Lost in the Middle: How Language Models Use Long Contexts},
  author  = {Liu, Nelson F. and Lin, Kevin and Hewitt, John and Paranjape, Ashwin and Bevilacqua, Michele and Petroni, Fabio and Liang, Percy},
  journal = {Transactions of the Association for Computational Linguistics},
  volume  = {12},
  pages   = {157--173},
  year    = {2024}
}

@article{xiao2023attentionsinks,
  title   = {Efficient Streaming Language Models with Attention Sinks},
  author  = {Xiao, Guangxuan and Tian, Yuandong and Chen, Beidi and Han, Song and Lewis, Mike},
  journal = {arXiv preprint arXiv:2309.17453},
  year    = {2023}
}

@article{su2024roformer,
  title   = {RoFormer: Enhanced Transformer with Rotary Position Embedding},
  author  = {Su, Jianlin and Ahmed, Murtadha and Lu, Yu and Pan, Shengfeng and Bo, Wen and Liu, Yunfeng},
  journal = {Neurocomputing},
  volume  = {568},
  pages   = {127063},
  year    = {2024}
}

@inproceedings{joachims2017unbiased,
  title     = {Unbiased Learning-to-Rank with Biased Feedback},
  author    = {Joachims, Thorsten and Swaminathan, Adith and Schnabel, Tobias},
  booktitle = {Proceedings of the Tenth ACM International Conference on Web Search and Data Mining},
  year      = {2017},
}

@inproceedings{ai2018unbiased,
  title     = {Unbiased Learning to Rank with Unbiased Propensity Estimation},
  author    = {Ai, Qingyao and Bi, Keping and Luo, Cheng and Guo, Jiafeng and Croft, W. Bruce},
  booktitle = {Proceedings of the 41st International ACM SIGIR Conference on Research and Development in Information Retrieval},
  year      = {2018}
  }

@inproceedings{hager2024ultr,
  title     = {Unbiased Learning to Rank Meets Reality: Lessons from Baidu's Large-Scale Search Dataset},
  author    = {Hager, Philipp and Deffayet, Romain and Renders, Jean-Michel and Zoeter, Onno and de Rijke, Maarten},
  booktitle = {Proceedings of the 47th International ACM SIGIR Conference on Research and Development in Information Retrieval},
  pages     = {1546--1556},
  year      = {2024}
}

@article{wang2023notfair,
  title   = {Large Language Models are not Fair Evaluators},
  author  = {Wang, Peiyi and Li, Lei and Chen, Liang and Cai, Zefan and Zhu, Dawei and Lin, Binghuai and Cao, Yunbo and Liu, Qi and Liu, Tianyu and Sui, Zhifang},
  journal = {arXiv preprint arXiv:2305.17926},
  year    = {2023}
}

@article{shi2025judgingjudges,
  title   = {Judging the Judges: A Systematic Study of Position Bias in LLM-as-a-Judge},
  author  = {Shi, Lin and Ma, Chiyu and Liang, Weicheng and Diao, Xingjian and Ma, Weining and Vosoughi, Sadegh},
  journal = {arXiv preprint arXiv:2406.07791},
  year    = {2025}
}

@article{li2024splitmerge,
  title   = {Split and Merge: Aligning Position Biases in LLM-based Evaluators},
  author  = {Li, Zongjie and Wang, Chaozheng and Ma, Pingchuan and Wu, Daoyuan and Wang, Shuai and Gao, Cuiyun and Liu, Yue},
  journal = {arXiv preprint arXiv:2310.01432},
  year    = {2024}
}

@article{sun2023chatgpt,
  title   = {Is ChatGPT Good at Search? Investigating Large Language Models as Re-Ranking Agents},
  author  = {Sun, Weiwei and Yan, Lingyong and Ma, Xinyu and Wang, Shuaiqiang and Ren, Pengjie and Chen, Zhumin and Yin, Dawei and Ren, Zhaochun},
  journal = {arXiv preprint arXiv:2304.09542},
  year    = {2023}
}

@inproceedings{chen2022denoising,
  title={Denoising self-attentive sequential recommendation},
  author={Chen, Huiyuan and Lin, Yusan and Pan, Menghai and Wang, Lan and Yeh, Chin-Chia Michael and Li, Xiaoting and Zheng, Yan and Wang, Fei and Yang, Hao},
  booktitle={Proceedings of the 16th ACM conference on recommender systems},
  year={2022}
}

@inproceedings{hou2024zeroshot,
  title     = {Large Language Models are Zero-Shot Rankers for Recommender Systems},
  author    = {Hou, Yupeng and Zhang, Junjie and Lin, Zihan and Lu, Hongyu and Xie, Ruobing and McAuley, Julian and Zhao, Wayne Xin},
  booktitle = {European Conference on Information Retrieval},
  pages     = {364--381},
  year      = {2024}
}

@inproceedings{he2020lightgcn,
  title     = {LightGCN: Simplifying and Powering Graph Convolution Network for Recommendation},
  author    = {He, Xiangnan and Deng, Kuan and Wang, Xiang and Li, Yan and Zhang, Yongdong and Wang, Meng},
  booktitle = {Proceedings of the 43rd International ACM SIGIR Conference on Research and Development in Information Retrieval},
  pages     = {639--648},
  year      = {2020}
}

@article{harper2015movielens,
  title   = {The MovieLens Datasets: History and Context},
  author  = {Harper, F. Maxwell and Konstan, Joseph A.},
  journal = {ACM Transactions on Interactive Intelligent Systems},
  volume  = {5},
  number  = {4},
  pages   = {1--19},
  year    = {2015}
}

@inproceedings{hou2024amazonreviews,
  title     = {Bridging Language and Items for Retrieval and Recommendation},
  author    = {Hou, Yupeng and Li, Jiacheng and He, Zhankui and Yan, An and Chen, Xiusi and McAuley, Julian},
  booktitle = {arXiv preprint arXiv:2403.03952},
  year      = {2024}
}

@article{chen2024premise,
  title   = {Premise order matters in reasoning with large language models},
  author  = {Chen, Xinyun and Chi, Ryan A and Wang, Xuezhi and Zhou, Denny},
  journal = {arXiv preprint arXiv:2402.08939},
  year    = {2024}
}

@article{egressy2026set,
  title={Set-llm: A permutation-invariant llm},
  author={Egressy, Beni and St{\"u}hmer, Jan},
  journal={Advances in Neural Information Processing Systems},
  volume={38},
  pages={62798--62834},
  year={2026}
}

@article{zhao2025simaug,
  title={SimAug: Enhancing recommendation with pretrained language models for dense and balanced data augmentation},
  author={Zhao, Yuying and Yang, Xiaodong and Chen, Huiyuan and Fan, Xiran and Wang, Yu and Cai, Yiwei and Derr, Tyler},
  journal={arXiv preprint arXiv:2505.01695},
  year={2025}
}

\appendix

\section{Additional Results and Diagnostics}
\label{app:additional}

\subsection{Dataset Construction Details}
\label{app:dataset}

\paragraph{Label construction.} Candidate labels come from temporally held-out future interactions rather than manual ranking annotations. Observed ratings map to graded relevance labels under a shared rule: ratings \(1\), \(2\), \(3\), and \(4/5\) become labels \(1\), \(2\), \(3\), and \(4\) respectively, so the top two rating levels are collapsed into a single highest-relevance label. Retrieved negatives and catalog filler items which sampled from the catalog to keep the candidate-set size fixed, receive label \(0\). Label \(0\) therefore denotes a non-positive candidate under the evaluation split, while labels above \(0\) denote observed future interactions with graded relevance.

Inserting the held-out positives directly is necessary in the Amazon domains rather than a convenience. Requiring LightGCN to retrieve the held-out future item---the \texttt{REQPOS=true} condition---would discard most otherwise usable evaluation instances, because first-stage Recall@100 is only about \(1.3\%\) on both Amazon domains (Table~\ref{tab:dataset_stats}). We therefore use \texttt{REQPOS=false} throughout. Because the positive is inserted at a random position rather than at a retrieval-determined one, its input position carries no information about relevance, which is a precondition for attributing rank changes to presentation order alone.

\paragraph{Ground-truth ranking and ties.} The ground-truth ranking is induced by decreasing relevance label. Candidates with equal labels are treated as ties, and the LambdaRank objective~\cite{burges2010ranknet, wang2018lambdaloss} is applied only to item pairs with unequal labels. Table~\ref{tab:dataset_stats} reports raw dataset sizes, filtering thresholds, split sizes, history windows, and first-stage retrieval recall for all three domains.

\begin{table*}[t]
\centering
\caption{%
\textbf{Dataset construction statistics.}
We report raw dataset sizes, filtering parameters, post-split sample counts, history lengths, and first-stage retrieval recall. First-stage recall was not logged for the MovieLens build and is omitted rather than estimated.}
\label{tab:dataset_stats}

\setlength{\tabcolsep}{3pt}
\renewcommand{\arraystretch}{0.95}

\resizebox{\textwidth}{!}{%
\begin{tabular}{lrrrrrrrccc}
\toprule
Dataset
& Raw inter.
& Raw items
& Hist.
& Min int.
& Train
& Val
& Test
& Recall@10
& Recall@50
& Recall@100 \\
\midrule
MovieLens-32M
& 32.0M
& 87.6K
& 20
& 50
& 9,998
& 9,996
& 9,997
& --
& --
& -- \\

Amazon Books
& 29.5M
& 4.45M
& 20
& 20
& 4,084
& 4,081
& 4,087
& 0.0024
& 0.0085
& 0.0133 \\

Amazon Fashion
& 2.50M
& 826K
& 5
& 8
& 2,003
& 1,974
& 2,070
& 0.0027
& 0.0086
& 0.0126 \\
\bottomrule
\end{tabular}%
}

\vspace{0.25em}
{\footnotesize
Amazon domains use \texttt{REQPOS=false}: requiring the held-out future positive to appear in the LightGCN retrieval set would remove most usable samples because first-stage recall is very low. Fashion uses a shorter history window and a lower minimum-interaction threshold to retain enough eligible users.
}
\end{table*}

\newtcolorbox{promptbox}[1]{%
  colback=gray!4,
  colframe=black!55,
  fonttitle=\bfseries,
  title={#1},
  boxrule=0.5pt,
  left=3pt,right=3pt,top=3pt,bottom=3pt,
  before skip=6pt,after skip=6pt,
  breakable,
  enhanced
}

\subsection{Prompt-Level Mitigation Does Not Close the Attack Surface}
\label{sec:prompt-mitigation}

A natural first reaction to the position-bias attack is to simply \emph{instruct} the
reranker to ignore candidate order. We test whether this works. Holding the \emph{same}
trained causal reranker fixed (same weights, causal attention, standard RoPE), we change
\emph{only} the eval-time instruction and re-run both probes---the position scan
(\textsc{curve\_range}) and the budget-$R$ permutation attacker (\textsc{\(\mathrm{promo}@5\)}).
If a prompt genuinely removes the bias, its metrics should collapse toward $0$, matching
the architectural / consistency-trained defenses; if they stay at the baseline, the bias
is \emph{mechanistic} rather than \emph{instructional}.

All prompts share the listwise template below; only the \textsc{Instruction} slot varies as shown in Table~\ref{tab:prompt-wordings}.
Candidate and history formatting are byte-identical across variants, so wording is the
sole moving part. 

\begin{promptbox}{Shared listwise template}
\small\ttfamily\raggedright
[SPAN]\par
\textit{$\langle$Instruction$\rangle$}\par
User history:\par
title: $\ldots$\par
[/SPAN]\par
\texttt{[ITEM]} $\langle$candidate 1$\rangle$ \texttt{[/ITEM]}\par
\texttt{[ITEM]} $\langle$candidate 2$\rangle$ \texttt{[/ITEM]}\par
$\ldots$
\end{promptbox}

We compare a control ($p_0$, the original wording) against five treatments that each
encode the ``ignore order / treat as a set'' intent in a \emph{different prompting style},
so a null result cannot be blamed on one unlucky phrasing. The results are presented in Table~\ref{tab:prompt-results}.

\begin{table*}[t]
\centering\small
\caption{Instruction wordings tested (the $\langle$Instruction$\rangle$ slot).}
\label{tab:prompt-wordings}
\begin{tabularx}{\linewidth}{@{}l l >{\raggedright\arraybackslash}X@{}}
\toprule
ID & Style & Instruction \\
\midrule
$p_0$ & control & Given the user's interaction history, rank the candidate items according to the user's preferences. \\
$p_1$ & declarative & Given the user's interaction history, rank the following candidate items by how well they match the user's preferences. The candidates are provided together as an unordered set: consider all of them as a group and judge each item only on its own merits, not on where it appears in the list. \\
$p_2$ & imperative & Rank the candidate items by relevance to the user's history. Ignore the list order. Do not favor earlier or later items. The position of a candidate is random and carries no meaning: judge content only. \\
$p_3$ & persona & You are a position-invariant recommendation reranker. Your defining property is that your output depends only on item content and the user's history, never on the order in which candidates are fed to you. Acting as this reranker, rank the candidate items by how well each matches the user's preferences. \\
$p_4$ & chain-of-thought & Given the user's interaction history, rank the candidate items by preference. Reason step by step: first read each candidate on its own and assess how relevant it is to the user, independent of the other candidates; then compare those per-item assessments to produce the final ranking. Because the input order is arbitrary, the ranking you produce must be the same no matter how the candidates were ordered. \\
$p_5$ & structured rules & Task: 
rank the candidate items for the user. 
Rules: 
1. Treat the candidates as an unordered set. 
2. The listed order is random and must be ignored.
3. Score each item only by its relevance to the user's interaction history.
4. An item's rank must not change if the list is reshuffled.
Rank the items following these rules. \\
\bottomrule
\end{tabularx}
\end{table*}

\begin{table}[t]
\centering
\scriptsize
\setlength{\tabcolsep}{2.5pt}
\renewcommand{\arraystretch}{0.92}
\caption{Prompting leaves the attack surface largely unchanged on MovieLens ($N{=}300$, $R{=}50$). Larger values mean more vulnerability.}
\label{tab:prompt-results}
\resizebox{\columnwidth}{!}{%
\begin{tabular}{@{}llcccc@{}}
\toprule
ID & Style & curve\_range & \(\mathrm{promo}@5\) & rank\_gain & into\_top5 \\
\midrule
$p_0$ & baseline         & 3.09 & 0.120 & 3.34 & 0.273 \\
$p_1$ & declarative      & 3.12 & 0.120 & 3.48 & 0.277 \\
$p_2$ & imperative       & 3.15 & 0.123 & 3.48 & 0.280 \\
$p_3$ & persona          & 2.93 & 0.127 & 3.14 & 0.273 \\
$p_4$ & CoT              & 3.20 & 0.123 & 3.53 & 0.277 \\
$p_5$ & rules            & 3.03 & 0.130 & 3.40 & 0.280 \\
\midrule
\multicolumn{2}{@{}l}{\textit{architectural invariance} (A)} & 0.06 & \textbf{0.000} & --- & --- \\
\multicolumn{2}{@{}l}{\textit{consistency-KL} (B2)}          & ---  & \textbf{0.007} & --- & --- \\
\bottomrule
\end{tabular}%
}
\end{table}

\paragraph{Finding.} Prompting is largely inert. Across all five styles, \textsc{\(\mathrm{promo}@5\)} stays around $0.12$--$0.13$ and \textsc{curve\_range} around $2.93$--$3.20$, essentially matching the control. By contrast, the architectural (A) and training-time (B2) defenses drive \textsc{\(\mathrm{promo}@5\)} to near zero. This suggests the position bias is mechanistic rather than instructional: it cannot be removed by prompting alone.

\subsection{Position vs. Content Diagnostic}
\label{app:position_content}

As a diagnostic on the MovieLens anchor, we compare the order-only lever with a simple content-edit lever. We cross two factors: position, either original or favorable, and content, either original or edited by injecting the user's top history genre into \(c^\ast\), as shown in Table~\ref{tab:position_content}. Repositioning \(c^\ast\) improves its rank by \(1.86\) positions on average, whereas this content edit improves it by \(0.33\) positions, with negligible interaction between the two effects. This suggests that candidate position can act as a strong content-free lever in this setting. We report this diagnostic to separate positional effects from content effects, not as a comprehensive comparison against optimized content attacks.

\begin{table}[t]
\centering
\caption{
\textbf{Position-vs.-content diagnostic on the MovieLens anchor.}
Rank improvement from repositioning a label-\(0\) target versus applying a simple relevance-signal content edit. Higher gain means stronger promotion.
}
\label{tab:position_content}
\small
\begin{tabular}{lc}
\toprule
Intervention & Mean rank gain \\
\midrule
Favorable position only & 1.86 \\
Content edit only & 0.33 \\
Interaction & negligible \\
\bottomrule
\end{tabular}
\end{table}

\subsection{Additional Defense Diagnostics}
\label{app:defense_details}

Table~\ref{tab:anchor_defense_full} reports the full anchor defense matrix on MovieLens, including mask-only invariance, position-only invariance, order-augmented training, permutation-consistency regularization, test-time averaging, and pointwise scoring. Table~\ref{tab:defense_cross_domain_full} reports the corresponding cross-domain stability diagnostics, including Kendall's \(\tau\) and label-\(0\) position-curve range.

\begin{table*}[t]
\centering
\caption{
\textbf{Full anchor defense matrix on MovieLens.}
This table reports all defense variants evaluated on the Llama-3.2-3B anchor. Lower \(\mathrm{promo}@5\) is better. Exact indicates whether the method is invariant by construction.
}
\label{tab:anchor_defense_full}
\small
\begin{tabular}{lcccc}
\toprule
Defense / model & nDCG@10 & \(\mathrm{promo}@5\) & Cost & Exact \\
\midrule
Baseline causal
& 0.829 & 0.120 & \(1\times\) & No \\

A: Architectural invariance
& 0.784 & \textbf{0.000} & \(1\times\) & Yes \\

A1: Mask-only invariance
& 0.783 & 0.070 & \(1\times\) & No \\

A2: Position-only invariance
& 0.769 & 0.090 & \(1\times\) & No \\

B1: Order-augmented training
& \textbf{0.849} & 0.117 & \(1\times\) & No \\

B2: Permutation-consistency KL
& 0.777 & 0.007 & \(1\times\) & Approx. \\

C: Test-time averaging \((P=20)\)
& 0.848 & \(\approx 0\) & \(20\times\) & No \\

Pointwise reference
& 0.763 & \textbf{0.000} & \(K\times\) & Yes \\
\bottomrule
\end{tabular}

\vspace{0.25em}
\footnotesize{
For test-time averaging, \(\mathrm{promo}@5\approx 0\) is inferred from averaged-score behavior rather than re-running the full adversarial search against the averaged scorer.
}
\end{table*}

\begin{table*}[t]
\centering
\caption{
\textbf{Full cross-domain defense diagnostics.}
In addition to ranking quality and manufactured exposure, we report permutation stability \(\tau\) and label-\(0\) position-curve range. Higher \(\tau\) and lower range indicate greater order stability.
}
\label{tab:defense_cross_domain_full}
\small
\begin{tabular}{llccccc}
\toprule
Domain & Defense / model & nDCG@10 & \(\mathrm{promo}@5\) & \(\tau\) & Range & Exact \\
\midrule
MovieLens
& Baseline causal
& 0.829 & 0.120 & 0.845 & 3.09 & No \\

MovieLens
& Architectural invariance
& 0.784 & 0.000 & 0.992 & 0.06 & Yes \\

MovieLens
& Permutation-consistency KL
& 0.777 & 0.007 & 0.983 & 0.21 & Approx. \\

MovieLens
& Pointwise reference
& 0.763 & 0.000 & 1.000 & 0.00 & Yes \\

\midrule
Amazon Books
& Baseline causal
& 0.618 & 0.480 & 0.563 & 6.62 & No \\

Amazon Books
& Architectural invariance
& 0.510 & 0.000 & 0.990 & 0.06 & Yes \\

Amazon Books
& Permutation-consistency KL
& 0.552 & 0.067 & 0.912 & 1.16 & Approx. \\

Amazon Books
& Pointwise reference
& -- & 0.000 & 1.000 & 0.00 & Yes \\

\midrule
Amazon Fashion
& Baseline causal
& 0.449 & 0.297 & 0.661 & 4.12 & No \\

Amazon Fashion
& Architectural invariance
& 0.424 & 0.003 & 0.984 & 0.06 & Yes \\

Amazon Fashion
& Permutation-consistency KL
& 0.424 & 0.047 & 0.934 & 0.65 & Approx. \\

Amazon Fashion
& Pointwise reference
& -- & 0.000 & 1.000 & 0.00 & Yes \\
\bottomrule
\end{tabular}
\end{table*}

\subsection{Full Cross-Model Sweep}
\label{app:cross_model}

Table~\ref{tab:cross_model_domain_full} reports the complete cross-model and cross-domain sweep corresponding to the scale and domain analysis in Section~\ref{sec:general}. The main text summarizes these results by domain-level ranges; here we provide the full per-model values.

\begin{table*}[t]
\centering
\caption{
\textbf{Full cross-model and cross-domain causal sweep.}
For each causal reranker, we report standard ranking quality using nDCG@10 and manufactured exposure using \(\mathrm{promo}@5\) at attack budget \(R=50\). Across the evaluated models, \(\mathrm{promo}@5\) does not monotonically decrease with parameter count; domain identity shifts the attack surface more strongly than scale. The final row reports the position-invariant reference on the anchor model.
}
\label{tab:cross_model_domain_full}
\small
\begin{tabular}{lcc cc cc}
\toprule
& \multicolumn{2}{c}{MovieLens}
& \multicolumn{2}{c}{Amazon Books}
& \multicolumn{2}{c}{Amazon Fashion} \\
\cmidrule(lr){2-3}
\cmidrule(lr){4-5}
\cmidrule(lr){6-7}
Model
& nDCG@10 & \(\mathrm{promo}@5\)
& nDCG@10 & \(\mathrm{promo}@5\)
& nDCG@10 & \(\mathrm{promo}@5\) \\
\midrule
Qwen3-0.6B
& 0.791 & 0.130
& 0.534 & 0.497
& 0.398 & 0.353 \\

Qwen3-1.7B
& 0.751 & 0.100
& 0.565 & 0.427
& 0.425 & 0.227 \\

Qwen3-4B
& 0.798 & 0.100
& 0.597 & 0.497
& 0.432 & 0.387 \\

Qwen3-8B
& 0.817 & 0.097
& 0.616 & 0.570
& 0.435 & 0.317 \\

Qwen3-14B
& 0.829 & 0.140
& 0.633 & 0.477
& 0.433 & 0.273 \\

Llama-3.1-8B
& 0.818 & 0.113
& 0.610 & 0.497
& 0.418 & 0.270 \\

Mistral-7B
& \textbf{0.865} & 0.147
& \textbf{0.646} & 0.540
& 0.448 & 0.267 \\

Llama-3.2-3B
& 0.829 & 0.120
& 0.618 & 0.480
& \textbf{0.449} & 0.297 \\

\midrule
Position-invariant reference
& 0.784 & \textbf{0.000}
& 0.510 & \textbf{0.000}
& 0.424 & \textbf{0.003} \\
\bottomrule
\end{tabular}
\end{table*}

\subsection{Audit-Correlation Robustness}
\label{app:audit_robustness}

Table~\ref{tab:tau_promo_robustness} reports robustness checks for the relationship between ordinary permutation stability and manufactured exposure. The negative relationship between Kendall's \(\tau\) and \(\mathrm{promo}@5\) remains strong under leave-one-domain-out checks and after controlling for domain identity.

\begin{table*}[t]
\centering
\caption{
\textbf{Audit-correlation robustness checks.}
We correlate causal-reranker permutation stability \(\tau\) with manufactured exposure \(\mathrm{promo}@5\). The relationship remains strong after controlling for domain identity and under leave-one-domain-out checks.
}
\label{tab:tau_promo_robustness}
\small
\begin{tabular}{lrrrrr}
\toprule
Analysis & \(n\) & Pearson \(r\) & Pearson \(p\) & Spearman \(\rho\) & Spearman \(p\) \\
\midrule
All 24 cells
& 24 & -0.974 & \(<10^{-4}\) & -0.969 & \(<10^{-4}\) \\

Amazon Books only
& 8 & -0.857 & 0.0066 & -0.683 & 0.0618 \\

Amazon Fashion only
& 8 & -0.812 & 0.0143 & -0.786 & 0.0208 \\

MovieLens only
& 8 & -0.541 & 0.1665 & -0.683 & 0.0621 \\

Leave out Amazon Books
& 16 & -0.964 & \(<10^{-4}\) & -0.935 & \(<10^{-4}\) \\

Leave out Amazon Fashion
& 16 & -0.993 & \(<10^{-4}\) & -0.923 & \(<10^{-4}\) \\

Leave out MovieLens
& 16 & -0.971 & \(<10^{-4}\) & -0.935 & \(<10^{-4}\) \\

Partial corr., control domain
& 24 & -0.776 & \(8\times10^{-6}\) & -- & -- \\
\bottomrule
\end{tabular}
\end{table*}

\subsection{Attack-Budget Values}
\label{app:budget_values}

Table~\ref{tab:budget_curve_values} provides the numerical values behind the attack-budget curve in Figure~\ref{fig:budget}. The causal reranker becomes more exploitable as the attacker tries more candidate orderings, while construction-based invariant references remain at zero.

\begin{table*}[t]
\centering
\caption{\textbf{Attack-budget sensitivity on the MovieLens anchor.} \(\mathrm{promo}@5\) is reported for label-\(0\) targets, from the same \(R_{\max}=50\) reference run as Table~\ref{tab:mechanism_comparison}.}
\label{tab:budget_curve_values}
\small
\begin{tabular}{lccccc}
\toprule
Model & \(R=1\) & \(R=5\) & \(R=10\) & \(R=20\) & \(R=50\) \\
\midrule
Causal Llama-3B
& 0.023 & 0.067 & 0.090 & 0.103 & 0.120 \\
T5 encoder scorer
& 0.007 & 0.030 & 0.040 & 0.047 & 0.057 \\
Architectural invariance
& 0.000 & 0.000 & 0.000 & 0.000 & 0.000 \\
Pointwise reference
& 0.000 & 0.000 & 0.000 & 0.000 & 0.000 \\
\bottomrule
\end{tabular}
\end{table*}

\end{document}